\documentclass[journal=jacsat,manuscript=article, layout=twocolumn]{achemso}

\usepackage{chemformula} 
\usepackage[T1]{fontenc} 
\usepackage{miller}
\usepackage{hyperref}

\author{Junghwa Kim}
\altaffiliation{These authors contributed equally to this work.}
\email{kjunghwa@mit.edu}
\author{Colin Gilgenbach}
\altaffiliation{These authors contributed equally to this work.}
\author{Aaditya Bhat}
\author{James LeBeau}
\email{lebeau@mit.edu}
\affiliation[MIT]
{Department of Materials Science and Engineering, Massachusetts Institute of Technology, Cambridge, 02139, Massachusetts, United States}

\title[An \textsf{achemso} demo]
    {Quantifying Implantation Induced Damage and Point Defects with Multislice Electron Ptychography}

\abbreviations{STEM, SiC}

\begin{document}




\begin{abstract}

Here, we use multislice electron ptychography to quantify damage introduced by ion implantation of Er into 4H-SiC. Comparing reconstructed volumes from experiment (each 2,000 nm$^{3}$) along the implantation direction, the crystal damage is quantified and compared to pristine SiC. Using simulations, we establish that the implantation-induced static displacements limit both Er dopant and silicon vacancy detection. The corresponding damage in the experiment is found to occur up a depth of 100 nm and significantly deeper than expected from implantation simulations, ignoring crystallography. Beyond this depth, we show that silicon vacancies can be identified within the sampled volume and used to measure their local strain. Overall, these results underscore the power of multislice electron ptychography to quantify the impacts of implantation and as a tool to help guide electronic device process optimization.
\end{abstract}




Controlled doping of semiconductors can be achieved through implantation, where ionized dopants are accelerated to an energy of tens to thousands of keV towards a target material.\cite{ionimplantation} Upon reaching the target material, their kinetic energy is transferred via collisions that displace atoms and/or create vacancies in a collision cascade\cite{implantationdamage,Sibombard,sicdefect_implant}. While the resulting damage cascade degrades electronic, thermal, and optical properties\cite{imp_electric,findefect,thermaldegradation,thermalprocess}, subsequent thermal annealing is used to reduce/eliminate damage and vacancies, and to activate the implanted dopants\cite{postannealing}. Accurately predicting post-annealing device performance will require direct chemical and structural analysis of the \textit{initial} damage and point defect distribution, which often remains elusive.

Few experimental techniques provide information about both damage and point defects, specifically atomic displacements and vacancies. Methods such as ion channeling, X-ray diffraction, and infrared reflectance have been used to reveal structural changes, including amorphization\cite{rbc,amorphization}, implantation damage-induced strain fields\cite{xrd}, and crystallographic disorder\cite{ir,ir2}. These conventional approaches, however, provide averaged information from the analysis volume. Consequently, a detailed correlation between induced point defects and their surrounding structure is often lacking. Beyond structural analysis, secondary ion mass spectrometry (SIMS) is usually utilized for accurate and precise dopant concentration profiles as a function of implantation depth. When surface topography is rough, such as after high-temperature annealing or when the implanted species are spatially non-homogeneous, SIMS-measured concentration profiles become less reliable\cite{annealingroughness,roughness,sims}. Recently, local lattice parameter measurements using 4D-STEM have enabled mapping vacancy distributions with nanometer-scale resolution\cite{mills2023nanoscale}, but challenges remain in probing and identifying single-atom-scale defects. Therefore, a method that can directly capture detailed structural information and measure dopant-level point defect concentration would be an invaluable tool in next-generation semiconductor device development, where device geometry is becoming increasingly complex\cite{complex1,cmos}.

Although direct point defect identification has been successfully demonstrated with annular dark field (ADF) scanning transmission electron microscopy (STEM)\cite{akiyama,dopants_5nm,srtio3vac,ishikawa}, dynamical scattering severely limits its applicability and reliability due to sample thickness constraints (usually $<$10 nm)\cite{muller,dopants_2nm,dopants_5nm}. This generally precludes the ability to detect dopants at semiconductor-relevant concentrations within the limited analysis volume. Moreover, the atomic number contrast in ADF-STEM further limits vacancy detection, particularly for light elements, such as carbon or oxygen\cite{ovac,srtio3vac}. 

Contrasting with conventional imaging, multislice electron ptychography (MEP)\cite{multiptychography,ovac,interstitial,adaptive} has overcome the challenges of projection and dynamical scattering by iteratively solving for the sample potential as a series of slices, thus preserving three-dimensional sample information. The approach readily enables the reconstruction of volumes $>$2,000 nm$^{3}$, or approximately 10 nm (length) $\times$ 10 nm (width) $\times$ 40 nm (thickness)\cite{ovac,multiptychography,colin}. Moreover, with atomic number sensitivity ($\sim Z^{0.67}$) and spatial resolution limited by the thermal motion of the atoms \cite{multiptychography}, vacancies, interstitials, and dopants\cite{ovac,multiptychography,interstitial} can be measured. These capabilities highlight MEP's potential as a near-ideal tool to probe the effects of implantation in a material while simultaneously characterizing sample structure with extremely high spatial resolution.

To quantify as-implanted sample damage with MEP, silicon carbide (SiC) serves as an ideal test case with widely available, near-defect free single crystal substrates\cite{single_crystals}. Moreover, it is composed of heavier and lighter elements (as opposed to only Si), exists in a variety of polymorphs (e.g.~3C, 4H, 6H), and it is widely used in electronic applications\cite{sicpower,ganled}. Furthermore, SiC is promising host material for scalable and integrated quantum applications due to its ability form a variety of spin-active defects created via implantation, such as divacanies or transition metal-vacancy complexes \cite{variousdefects}. Among various SiC dopants to test with MEP, erbium (Er) offers large atomic differences with the host and is widely used in optical devices\cite{Er}.

In this Letter, we quantify damage in a 4H-SiC host resulting from Er ion implantation. This is achieved using MEP to reconstruct volumes at multiple locations along the implantation direction, each containing over 10$^5$ atoms. By analyzing the local atomic structure at each implantation depth, we directly show that the induced damage extends significantly further into the sample than predicted by implantation simulations that ignore crystallography. Further, we evaluate the sensitivity of this technique to Si vacancies (vac$_\textrm{Si}$) as a function of damage to locate them within the reconstructed volumes and measure their local, surrounding distortions. Overall, this study demonstrates the effectiveness of MEP in analyzing implantation damage.





\begin{figure}[htbp]
\centering
\includegraphics[width=3.1in]{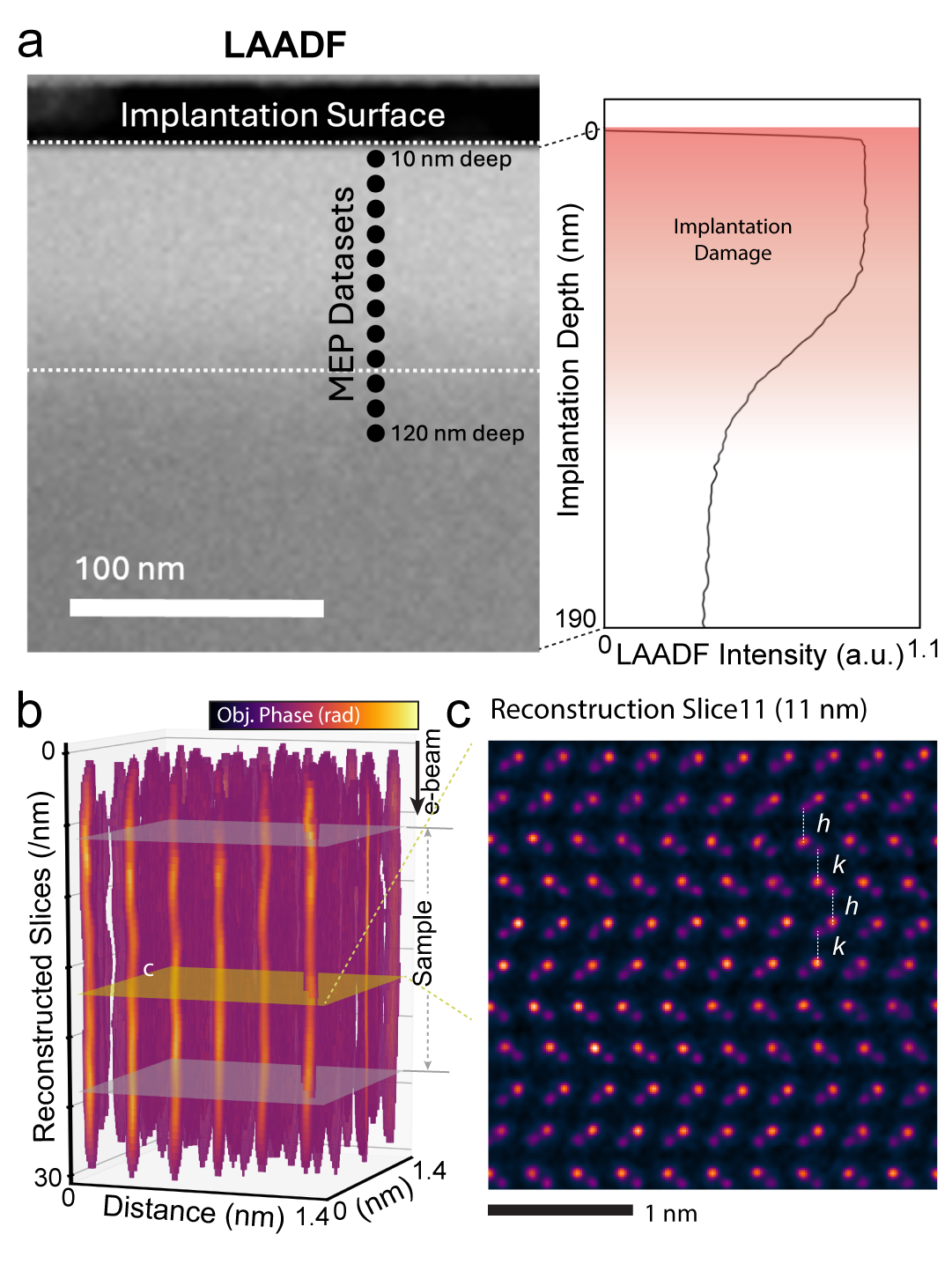}
\caption{(a) Cross-section LAADF-STEM image of Er-implanted 4H-SiC (left) and corresponding LAADF intensity profile (right). (b) 3D representation of the sample from the MEP reconstructed object slice stack. (c) A single reconstructed phase image from 11 nm below the TEM sample surface.}\label{ptychoreconstruct}
\end{figure}

A cross-sectional low-angle annular dark-field (LAADF) STEM overview of the Er implantated 4H-SiC is shown in Figure \ref{ptychoreconstruct}a. The highest LAADF intensity occurs at the implantation surface, which then decreases to at constant level at around 100 nm deep. This suggests significant crystallographic damage (static displacements) within this region \cite{phillips2012atomic}. Quantification of this intensity change is, however, challenging due to contributions from specimen thickness, surface contamination, dopants, strain, and implantation-induced displacements \cite{LAADF_Muller_oxygen,kumar2014dynamics}.

Instead, MEP is used to reconstruct sample volumes within this damaged layer at positions every 10 nm along the implantation direction, starting 10 nm from the surface and ending 120 nm deep, as indicated in Figure \ref{ptychoreconstruct}a. The reconstructed phase is proportional to the projected atomic potential, offering atomic number sensitivity and depth sensitivity.\cite{multiptychography,ovac} For example, the reconstructed phase clearly resolves both Si and C sublattices with three-dimensional information (Figure \ref{ptychoreconstruct}b-c) and with extreme spatial resolution (30 pm information limit, Figure S3)\cite{multiptychography}. Moreover, the non-equivalent \textit{h} (quasi-hexagonal) and \textit{k} (quasi-cubic) sites of 4H-SiC\cite{4HSiC_stacking} are readily distinguished, see Figure \ref{ptychoreconstruct}c. 

Because SiC is not readily wet-etched \cite{Pearton2007-xn}, removal of surface damage before imaging is impractical, other than minimizing through low-energy ion milling as applied here. Thus, distinguishing isolated defects from imperfections in the SiC sample surface prepared for electron microscopy remains challenging in conventional STEM. With MEP, however, the surface layers in the ptychographic reconstruction can be removed by subtracting surface slices from the volume. Using the standard deviation ($\sigma$) of each reconstructed phase slice (Figure S3), the near-surface regions are identified by an abrupt decrease in reconstructed phase contrast due to amorphization and the blurring of the vacuum and crystal layers.

\begin{figure}[htbp]
\centering
\includegraphics[width=3.1in]{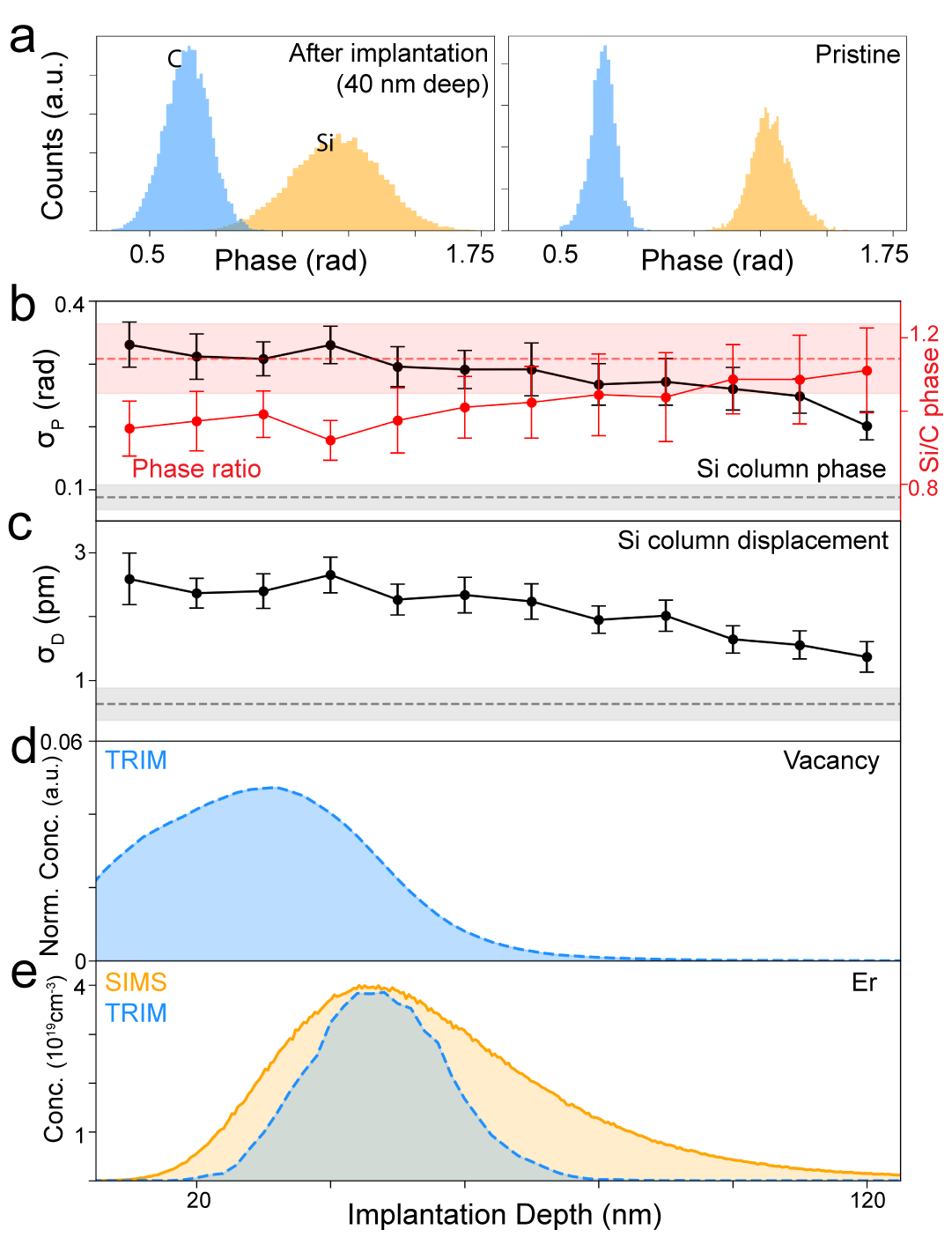}
\caption{(a) Phase distribution of Si and C sublattices in Er-implanted (left) and pristine SiC (right). (b) Si phase variation ($\sigma$\textsubscript{P}, black line) and Si/C phase ratio (red line), and (c) Si atom displacements ($\sigma$\textsubscript{D}) over implantation depth. Dashed lines and shaded areas in (b,c) are $\sigma$\textsubscript{P} (black) / peak ratio (red) and $\sigma$\textsubscript{D} measured in a pristine 4H-SiC sample. (d) TRIM calculated vacancy concentration profile due to Er ion implantation. (e) Depth profile of Er ion using SIMS (orange) and TRIM calculation (blue). }\label{depthdependent}
\end{figure}

The peak phase (maximum phase from a fitted peak function) at Si and C positions in the reconstructions are extracted for each slice and at each implantation depth, by fitting to a two-dimensional pseudo-Voigt function (see Supplementary Note S2, Figure S4). In the implanted sample (at an implantation depth of 40 nm, shown in Figure \ref{depthdependent}a, left panel), the Si and C peak phase distributions overlap, contrasting with the pristine, unimplanted sample volume, where Si consistently exhibits a significantly higher peak phase than C (Figure \ref{depthdependent}a, right panel). The implantation depth-dependent reconstructed phase distributions for Si and C are shown in Figure S5. With increasing implantation depth, the Si peak phase gradually increases, while the C phase remains largely constant. Consequently, the peak phase ratios of Si to C are 0.92$\pm$0.05 and 1.13$\pm$0.13 at 40 nm and 120 nm (a 23\% change), respectively (Figure \ref{depthdependent}b, red solid line), and the value at 120 nm depth exhibits a similar to that of the unimplanted sample (Si/C ratio 1.14$\pm$0.09, Figure \ref{depthdependent}b, dashed, horizontal line).


The reconstructed phase variation is analyzed as a function of implantation depth by measuring the standard deviation of the Si peak phase ($\sigma_\textrm{P}$). As shown in Figure \ref{depthdependent}b, $\sigma_\textrm{P}$ reaches its maximum and minimum at implantation depths of 40 and 120 nm, respectively. The trend is consistent with the standard deviation of the displacements of Si away from their column depth-averaged positions, $\sigma$\textsubscript{D}, Figure \ref{depthdependent}c. Up to an implantation depth of 40 nm, $\sigma$\textsubscript{D} is roughly constant, 2.53 $\pm$ 0.12 pm, and aligns with the damage profile from Transport of Ions in Matter (TRIM) calculations (Figure \ref{depthdependent}d) ignoring crystallography. Specifically, TRIM simulations using the experimental implantation conditions predict significant vacancy creation in the vicinity of the surface due to atomic Si and C displacements, with the highest probability at roughly 32 nm from the implantation surface and maximum penetration range up to $\sim$140 nm (Figure \ref{depthdependent}d). Simultaneously, the maximum Er ion concentration is on the order of 4 $\times$ 10$^\textrm{19}$ ions/cm$^\textrm{3}$ at an implantation depth of $\sim$45 nm (Figure \ref{depthdependent}e, dashed line). 

Over the next 80 nm of implantation depth, $\sigma_{\mathrm{D}}$ gradually decreases to 1.38 pm, contrasting with the damage extent predicted by TRIM calculations. This discrepancy can be explained by channeling along low-index zones, which increases implantation depth beyond that expected in an amorphous model\cite{impactofchanneling}. Moreover, at an implantation depth of 120 nm, $\sigma_{\mathrm{D}}$ and $\sigma_{\mathrm{P}}$ remain above that of a pristine SiC ($\sigma_{\mathrm{D}}$ = 0.95 $\pm$ 0.2 pm, $\sigma_{\mathrm{P}}$ = 0.088 $\pm$ 0.020 rad). Thus, even minute displacements induced by implantation can be measured using MEP.



\begin{figure}[htp]
\centering
\includegraphics[width=3.1in]{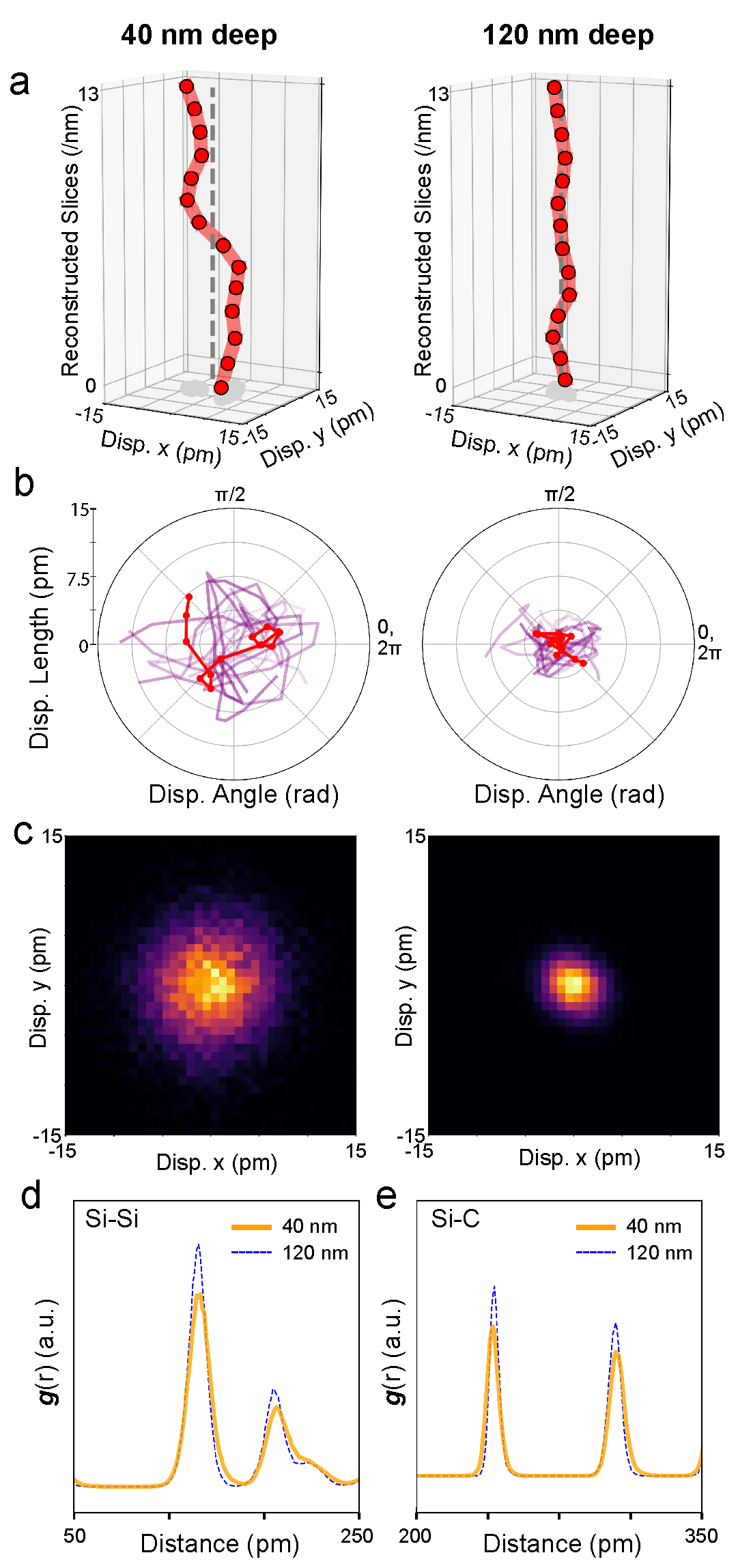}
\caption{(a-c) Atomic displacement mapping obtained at 40 nm (left) and 120 nm implantation depth (right). (a) 3D representations of randomly chosen Si columns, with gray dotted lines indicating the depth-averaged position of each column. (b) Polar plots for Si displacements perpendicular to the electron beam in 10 randomly selected atom columns (c) The full 2D histrograms of all of atomic displacements in the reconstruction. pPDF for (d) Si-Si and (e) Si-C distances at 40 nm (orange solid line) and 120 nm (blue dashed line) from the implantation surface.}\label{implantationdefect}
\end{figure}

Beyond the statistical average of damage within a reconstructed volume, the implantation-induced displacements can also be directly visualized \textit{along} the atom columns, as observed in Supplementary Videos 1 and 2 at implantation depths of 40 and 120 nm, respectively. Two randomly selected, representative Si atom columns and their displacements away from the column depth-average (dashed line) are shown in Figure \ref{implantationdefect}a. For the 40 nm deep case, Si exhibits significant \textit{xy}-plane displacements up to approximately 20 pm, while at 120 nm the maximum displacement is 10 pm. 

For the two implantation depths, the displacement magnitude and angle perpendicular to the electron beam for 10 randomly chosen atom columns are shown in Figure \ref{implantationdefect}b, and the 2D histogram of all projected displacements are shown in Figure \ref{implantationdefect}c). In each case, the displacement distributions are approximately isotropic, with significantly larger displacements occurring at 40 nm deep compared to 120 nm deep. Moreover, across all datasets (pristine, 40 nm, and 120 nm implantation depths), the C sublattice exhibits higher displacement magnitudes than Si (Figure S6, Table S1), likely due to the lower displacement threshold of C compared to Si\cite{meyer2012accurate}.

To gain quantitative insights into the local structure of as-implanted 4H-SiC, the projected pair distribution function (pPDF) is calculated\cite{Xiahan}. The pPDFs at the indicated implantation depths are shown in Figure \ref{implantationdefect}d, while the entire series is shown in Figure S7. The distances between first (NLN) and second Si nearest like-neighbors (2NLN) are slightly larger on average and exhibit higher scatter at an implantation depth of 40 nm ($\sigma_{NLN}$ = 30.34 pm and $\sigma_{2NLN} =$ 34.87 pm) compared to 120 nm ($\sigma_{NLN}$ = 24.32 pm and $\sigma_{2NLN} =$ 26.48 pm). The scatter of the nearest neighbor ($\sigma_{NN}$) distance, Si-C, is 28.29 pm at 40 nm or 20\% larger than at an implantation depth of 120 nm. 

While previous studies have demonstrated that severely damaged and amorphized SiC can exhibit radial distribution function peak at an interatomic distance of 145 pm\cite{edrdf,md_damage2}, corresponding to the equilibrium bond length of C-C, the pPDFs measured here (Figure S7b) indicate that the structure is not yet amorphized in the analyzed regions, with no evidence of C–C bond formation at this implantation dose. These measurements demonstrate the sensitivity of ptychography to quantify implantation-induced damage, which introduces measurable crystallographic disorder at the short- and medium-range length scales\cite{adp}. 

In addition to quantifying structural damage, the reconstructed phase can also be highly sensitive to point defects \cite{multiptychography,ovac}. For an ideal, undamaged crystal containing Er$_\textrm{Si}$ dopants, ptychography reconstructions from simulated 4D STEM datasets show that the peak phase increases by roughly 20\% at the defect position along the depth of the TEM sample (Figure S8). Simulating increasing levels of static displacements of atoms along an atom column (Figure S9), however, reveals that damage-induced displacements significantly broaden the peak phase due to blurring of the projected potential, and precludes direct defect identification when significant damage is present. The experiment further corroborates this, where large values of $\sigma_\textrm{P}$ occur near the implantation surface, Figure \ref{depthdependent}b, which then decreases with increasing implantation depth. Consequently, point defect detection in 4H-SiC is only feasible in regions with minimal implantation damage. We define such regions as those where the separation between Si and C phase is at least five times greater than$\sigma_\textrm{P,Si}$, minimizing the overlap between Si and C phases (Figure S10).

\begin{figure}[htbp]
\centering
\includegraphics[width=3.1in]{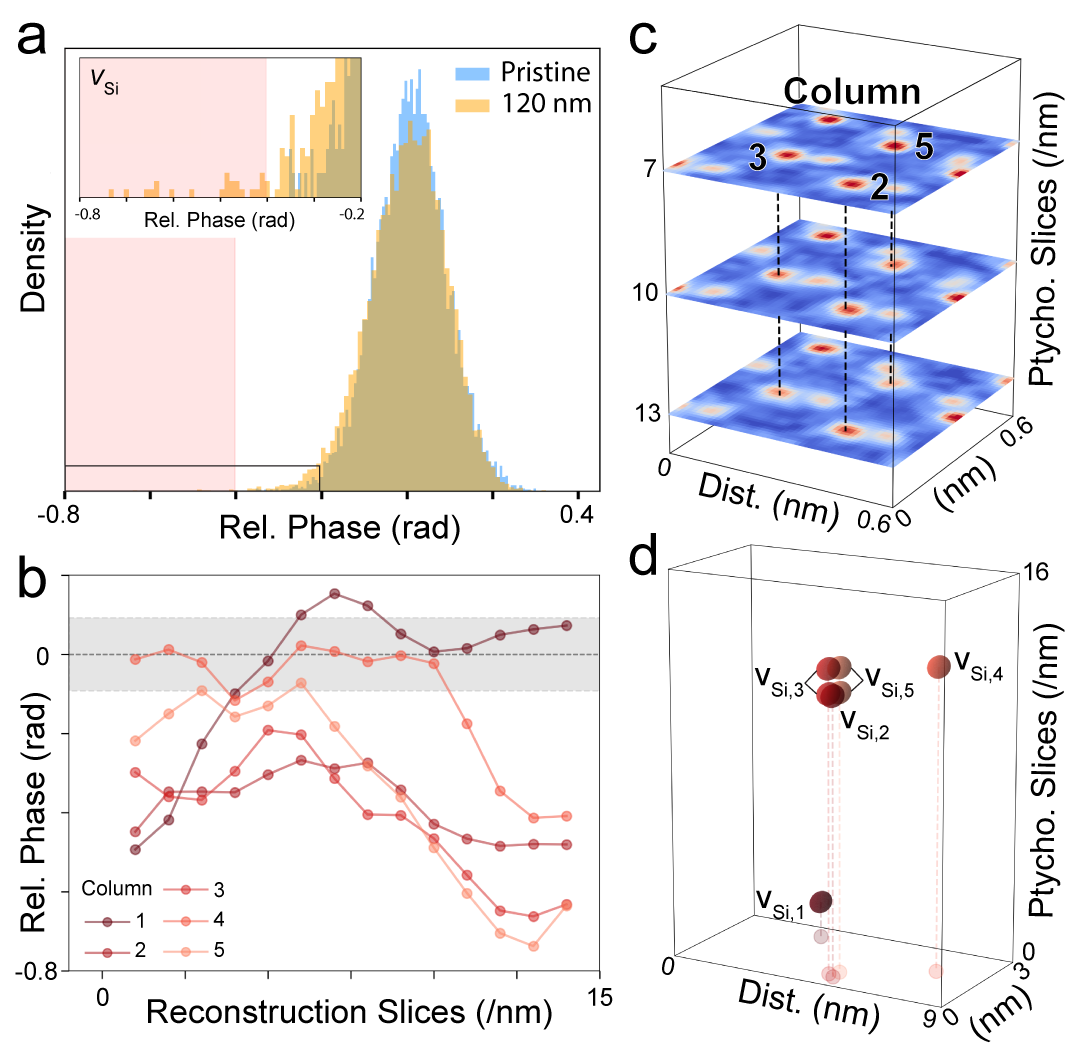}
\caption{(a) Si phase distributions at 120 nm implantation depth (orange) and from the pristine sample (blue). The inset highlights the left-hand tail of the distribution at 120 nm depth. The red-shaded area indicates Si phases below the pristine minimum. (b) Isolated atom columns potentially containing \ch{vac_{Si}} and their phase profiles over slices. Gray shaded regions represent Si phase within 1$\sigma$\textsubscript{P}. (c) Three representative slices (7, 10, 13) with defect-containing columns (2, 3, 5) as indicated. (d) 3D distribution of \ch{vac_{Si}} within the MEP volume (9 nm $\times$ 9 nm $\times$ 16 nm).}\label{distribution}
\end{figure}

Here, the aforementioned damage threshold is reached for depths $>$100 nm from the substrate surface, as indicated by the region outside the red-shaded region in Figure \ref{ptychoreconstruct}a. At 120 nm deep, for example, the Si phase histogram shows a pronounced left-side tail compared to the pristine SiC (Figure \ref{distribution}a). Simulated ptychography reconstructions with a similar displacement (\ch{$\sigma$_{D}} $\sim$1.4 pm) reveal that \ch{Er_{Si}} and \ch{vac_{Si}} (Figure S11) induce long tails on the right and left sides of the histogram, respectively. Thus, outliers with reduced reconstructed phase can be used to identify \ch{vac_{Si}}, while outliers with increased reconstructed phase can indicate \ch{Er_{Si}}. 

Although these outliers are linked to point defects, defining a robust cutoff to determine true outliers from statistical variation remains challenging due to dataset-specific offsets and the difficulty of establishing a suitable statistical model (Figure S12). As an estimate, a pristine sample reconstructed volume containing around 10,000 data points is used to derive the reconstructed object phase statistics empirically, Figure \ref{distribution}a. For consistent comparison across the datasets, the average phase of each dataset is offset to zero. 



Si sites with peak phase lower than the minimum Si phase of the pristine sample are classified as implantation-induced vacancies (red-shaded region in Figure \ref{distribution}a). While there are 17 points in the histogram matching this criterion, these points neighbor one another and are localized within five atom columns. This is validated by the peak phase line profile along these atom columns, shown in Figure \ref{distribution}b, where smooth variation is seen through the sample depth. In contrast, for the pristine sample data (different dataset), the lowest peak phase is confined to the sample surface (Figure S13), potentially due to surface roughness or vacancies introduced during TEM sample preparation/data acquisition.

While the vacancy containing atom columns selected from the implanted sample exhibit a similar Si phase at slice 7, significant phase reduction is observed in columns 3 and 5 at slice 13 nm, \textit{i.e.}~the presence of \ch{vac_{Si}}. Additionally, columns 3 and 5 show notably lower Si peak phase values (about -0.6 $\pm$ 0.04 rad below from the average Si phase), suggesting multiple vacancies are involved. Simulated reconstructions (Figure S11), for example, indicate that two vacancies within the depth resolution reduce the Si peak phase by around double that for a single vacancy. The minimum peak phase along the depth in Figure \ref{distribution}b provides the z-coordinate of the vacancy (or vacancies), making it possible to locate them within the reconstructed volume, Figure \ref{distribution}d. While clustering is observed within the volume, this can be anticipated by the trajectory of high energy ions scattering within the crystal. Given the reconstructed volume size ($\sim$ 1,300 nm$^3$) and number of atoms (10$^5$), this corresponds to vacancy concentration on the order of $\sim$100 ppm. 



In contrast to the vacancy-containing columns, the 120 nm deep reconstruction exhibits a single Si site with a significantly increased peak phase at a Si position. This is 13.4\% higher than the maximum phases observed in the unimplanted samples. Empirically, given that 10,000 points were analyzed in the unimplanted sample, this points to an estimated probability of 0.01\% or less of observing this peak phase from a pristine Si column, strongly suggesting the identification of a potential \ch{Er_{Si}} dopant (Figure S14). A single \ch{Er_{Si}} detection using MEP at a depth of 120 nm corresponds to approximately 1$\times$10\textsuperscript{18} cm$^{-3}$ ($\sim$10 ppm), with a range of 0.8 to 1.2$\times$10$^{18}$cm$^{-3}$ for a sample thickness 16 nm $\pm$ 2 nm. The secondary ion mass spectrometry (SIMS) concentration profile supports this finding, showing a concentration of Er around 2$\times$10\textsuperscript{18} cm$^{-3}$ at 120 nm depth (yellow line in Figure \ref{depthdependent}e).



\begin{figure}[htbp]
\centering
\includegraphics[width=3.1in]{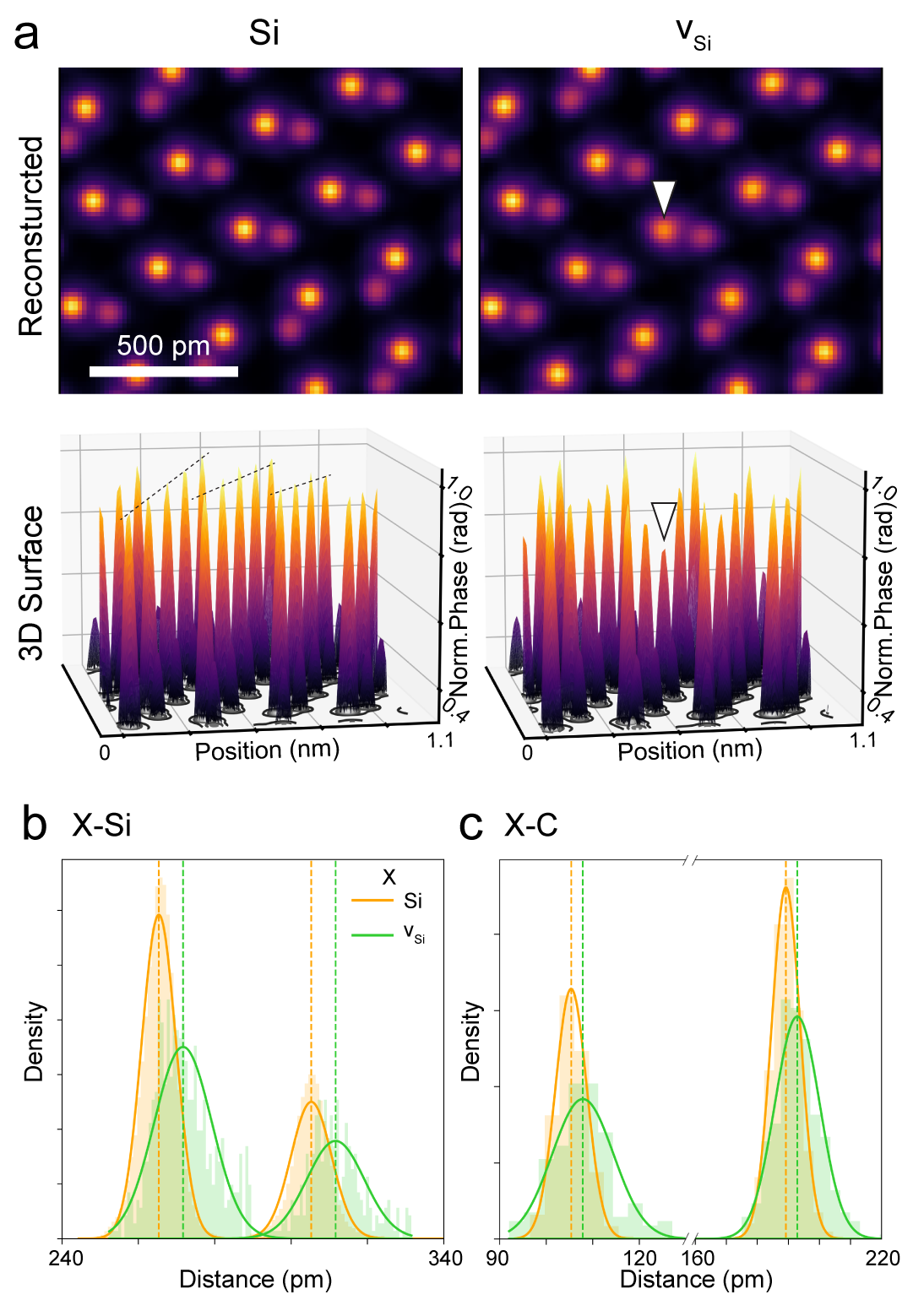}
\caption{\textbf (a) Class average phase patches for pristine Si and \ch{vac_{Si}} (left to right), with bottom panels showing 3D surface maps. (b,c) Distribution of bond lengths in the vicinity of Si (orange) and \ch{vac_{Si}} (green) with neighboring Si (b) and C (c) using multislice electron ptychography.}\label{localstructure}
\end{figure}

Beyond detecting individual point defects, ptychography also enables quantification of the local structure surrounding the point defects. With \ch{vac_{Si}} positions defined as above, non-defective Si positions are selected from those within one $\sigma_\mathrm{P}$ from the average Si phase. Using these positions, regional class averages for Si and \ch{vac_{Si}} are determined by extracting and averaging the 0.8 $\times$ 0.8 nm$^2$ regions around each and shown in Figure \ref{localstructure}a. Comparing the defect reconstructed phase to Si, \ch{vac_{Si}} exhibits a phase 0.25 rad that is lower. This phase difference corresponds to a 24$\%$ decrease versus the average Si column. The distances between the Si NLN at \ch{vac_{Si}} defects exhibit an expansion of 2.4 \% (with a standard error of 0.13 \%) to its nearest Si neighbor (Figure \ref{localstructure}b), which can be explained by outwards relaxation due to the localization of C dangling bonds\cite{distance}. For the C NLNs, around 2 \% expansion (with a standard error of 0.19 \%) is also observed at \ch{vac_{Si}} (Figure \ref{localstructure}c). This observation, coupled with the skewed distribution of the pPDF, shown in Figure \ref{implantationdefect}d,e, suggests that the crystal damage is attributable to the accumulation of vacancies originating from implanted ion collisions.




In summary, multislice electron ptychography enables three-dimensional quantification of crystal damage induced from implantation. Using outlier detection based on an empirical model determined using pristine 4H-SiC, the distribution of vac$_\textrm{Si}$ point defects can be identified. In addition, this approach also provides detailed local structure in the vicinity of point defects at concentrations relevant to semiconductor manufacturing. Overall, these findings address the long-standing need to characterize the structural consequences of ion implantation directly, \textit{e.g.} host crystal damage or local structure surrounding point defects, and can help overcome the limitations of other techniques to optimize device processing.

\begin{acknowledgement}

The authors thank support from the Air Force Office of Scientific Research (FA9550-22-1-0370). This work was performed with the assistance of MIT SuperCloud and was carried out in part through the use of MIT's Characterization.nano facilities. 

\end{acknowledgement}

\begin{suppinfo}

The Supporting Information is available free of charge.

Detailed experimental methods; supplementary notes and supplementary figures, including conventional STEM image and ptychographic reconstruction of erbium implanted 4H-SiC, atomic displacement and projected pair distribution function using MEP, the effect of atom displacement and point defects on reconstructed phase distribution using simulated ptychographic reconstruction, and potential \ch{Er_{Si}} and its 3D position in experiment using MEP (PDF).

\begin{itemize}
 \item Supplementary Video 1: A series of reconstructed slices at an implantation depth of 40 nm, with each slice having a thickness of 1 nm (.avi).
 \item Supplementary Video 2: A series of reconstructed slices at an implantation depth of 120 nm, with each slice having a thickness of 1 nm (.avi).
\end{itemize}

\end{suppinfo}

\bibliography{bibliography}

\end{document}